\documentclass[letter,longauth]{aa}
\pdfoutput=1  
\usepackage{natbib}
\usepackage[]{graphicx}
\usepackage{txfonts}

\def\pow#1#2{#1$\times$10$^{#2}$}
\def\micron{$\mu$m}
\def\kms{$\mathrm{km}\,\mathrm{s}^{-1}$}  
\def\psqcm{$\mathrm{cm}^{-2}$}
\def\pccm{$\mathrm{cm}^{-3}$}
\def\Tmb{$T_\mathrm{mb}$}  
\def\Tex{$T_\mathrm{ex}$}   
\def\HII{\hbox{\rm H\,{\sc ii}}\ }  
\def\vlsr{$V_\mathrm{lsr}$}  

\def\methanol{CH$_3$OH}

\begin{document}

\title{\textit{Herschel}/HIFI observations of spectrally resolved methylidyne signatures toward the high-mass star-forming core NGC\,6334I\thanks{\textit{Herschel} is an ESA space observatory with science instruments provided by European-led Principal Investigator consortia and with important participation from NASA.}}

\titlerunning{\textit{Herschel}/HIFI observations of CH toward NGC\,6334I}

\author{
M.H.D.~van der Wiel \inst{\ref{kapteyn},\ref{sron}} \and 
F.F.S.~van der Tak \inst{\ref{sron},\ref{kapteyn}} \and 
D.C.~Lis \inst{\ref{caltech}} \and 
T.~Bell \inst{\ref{caltech}} \and
E.A.~Bergin \inst{\ref{annarbor}} \and
C.~Comito \inst{\ref{bonn}} \and
M.~Emprechtinger \inst{\ref{caltech}} \and  
P.~Schilke \inst{\ref{bonn},\ref{koln}} \and
E.~Caux \inst{\ref{cesr},\ref{cesr2}} \and
C.~Ceccarelli \inst{\ref{laog},\ref{bordeaux},\ref{bordeaux2}} \and
A.~Baudry \inst{\ref{bordeaux},\ref{bordeaux2}} \and
P.F.~Goldsmith \inst{\ref{jpl}} \and
E.~Herbst \inst{\ref{ohio}} \and
W.~Langer \inst{\ref{jpl}} \and
S.~Lord \inst{\ref{ipac}} \and
D.~Neufeld \inst{\ref{hopkins}} \and
J.~Pearson \inst{\ref{jpl}} \and
T.~Phillips \inst{\ref{caltech}} \and
R.~Rolffs \inst{\ref{koln},\ref{bonn}} \and  
H.~Yorke \inst{\ref{jpl}} \and
A.~Bacmann \inst{\ref{laog},\ref{bordeaux},\ref{bordeaux2}} \and
M.~Benedettini \inst{\ref{ifsi}} \and
G.A.~Blake \inst{\ref{caltech}} \and
A.~Boogert \inst{\ref{ipac}} \and
S.~Bottinelli \inst{\ref{cesr},\ref{cesr2}} \and
S.~Cabrit \inst{\ref{lerma}} \and
P.~Caselli \inst{\ref{leeds}} \and
A.~Castets \inst{\ref{laog},\ref{bordeaux},\ref{bordeaux2}} \and
J.~Cernicharo \inst{\ref{cab}} \and
C.~Codella \inst{\ref{arcetri}} \and
A.~Coutens \inst{\ref{cesr},\ref{cesr2}} \and
N.~Crimier \inst{\ref{laog},\ref{cab}} \and
K.~Demyk \inst{\ref{cesr},\ref{cesr2}} \and
C.~Dominik \inst{\ref{uva},\ref{nijmegen}} \and
P.~Encrenaz \inst{\ref{lerma}} \and
E.~Falgarone \inst{\ref{lerma}} \and
A.~Fuente \inst{\ref{oan}} \and
M.~Gerin \inst{\ref{lerma}} \and
F.~Helmich \inst{\ref{sron}} \and
P.~Hennebelle \inst{\ref{lerma}} \and
T.~Henning \inst{\ref{heidelberg}} \and
P.~Hily-Blant \inst{\ref{laog}} \and
T.~Jacq \inst{\ref{bordeaux},\ref{bordeaux2}} \and
C.~Kahane \inst{\ref{laog}} \and
M.~Kama \inst{\ref{uva}} \and
A.~Klotz \inst{\ref{cesr},\ref{cesr2}} \and
B.~Lefloch \inst{\ref{laog}} \and
A.~Lorenzani \inst{\ref{arcetri}} \and
S.~Maret \inst{\ref{laog}} \and
G.~Melnick \inst{\ref{cfa}} \and
B.~Nisini \inst{\ref{oar}} \and
S.~Pacheco \inst{\ref{laog}} \and
L.~Pagani \inst{\ref{lerma}} \and
B.~Parise \inst{\ref{bonn}} \and
M.~Salez \inst{\ref{lerma}} \and
P.~Saraceno \inst{\ref{ifsi}} \and
K.~Schuster \inst{\ref{iram}} \and
A.G.G.M.~Tielens \inst{\ref{leiden}} \and
C.~Vastel \inst{\ref{cesr},\ref{cesr2}} \and
S.~Viti \inst{\ref{ucl}} \and
V.~Wakelam \inst{\ref{bordeaux},\ref{bordeaux2}} \and
A.~Walters \inst{\ref{cesr},\ref{cesr2}} \and
F.~Wyrowski \inst{\ref{bonn}} \and
%
K.~Edwards \inst{\ref{waterloo}} \and
J.~Zmuidzinas \inst{\ref{caltech}} \and
P.~Morris \inst{\ref{caltech}} \and
L.A.~Samoska \inst{\ref{jpl}} \and
D.~Teyssier \inst{\ref{esac}}
}

   \institute{
Kapteyn Astronomical Institute, University of Groningen, P.O.~Box 800, 9700AV, Groningen, The Netherlands; \texttt{wiel@astro.rug.nl}
\label{kapteyn}
\and SRON Netherlands Institute for Space Research, Groningen, NL
\label{sron}
\and California Institute of Technology, Pasadena, USA
\label{caltech}
\and Department of Astronomy, University of Michigan, Ann Arbor, USA
\label{annarbor}
\and Max-Planck-Institut f\"{u}r Radioastronomie, Bonn, Germany
\label{bonn}
\and Physikalisches Institut, Universit\"{a}t zu K\"{o}ln, K\"{o}ln, Germany
\label{koln}
\and Centre d'Etude Spatiale des Rayonnements, Universit\'e Paul Sabatier, Toulouse, France
\label{cesr}
\and CNRS/INSU, UMR 5187, Toulouse, France
\label{cesr2}
\and Laboratoire d'Astrophysique de Grenoble, UMR 5571-CNRS, Universit\'e Joseph Fourier, Grenoble, France
\label{laog}
\and Universit\'{e} de Bordeaux, Laboratoire d'Astrophysique de Bordeaux, Floirac, France
\label{bordeaux}
\and CNRS/INSU, UMR 5804, Floirac cedex, France
\label{bordeaux2}
\and Jet Propulsion Laboratory,  Caltech, Pasadena, CA 91109, USA\label{jpl}
\and Ohio State University, Columbus, OH, USA
\label{ohio}
\and INAF - Istituto di Fisica dello Spazio Interplanetario, Roma, Italy
\label{ifsi}
\and Infared Processing and Analysis Center,  Caltech, Pasadena, USA
\label{ipac}
\and Johns Hopkins University, Baltimore MD, USA
\label{hopkins}
\and Laboratoire d'Etudes du Rayonnement et de la Mati\`ere en Astrophysique, UMR 8112  CNRS/INSU, OP, ENS, UPMC, UCP, Paris, France
\label{lerma}
\and School of Physics and Astronomy, University of Leeds, Leeds UK
\label{leeds}
\and Centro de Astrobiolog\`{\i}a, CSIC-INTA, Madrid, Spain
\label{cab}
\and INAF Osservatorio Astrofisico di Arcetri, Florence Italy
\label{arcetri}
\and Astronomical Institute `Anton Pannekoek', University of Amsterdam, Amsterdam, The Netherlands
\label{uva}
\and Department of Astrophysics/IMAPP, Radboud University Nijmegen, Nijmegen, The Netherlands
\label{nijmegen}
\and IGN Observatorio Astron\'{o}mico Nacional, Alcal\'{a} de Henares, Spain\label{oan}
\and Max-Planck-Institut f\"ur Astronomie, Heidelberg, Germany
\label{heidelberg}
\and \mbox{Harvard-Smithsonian Center for Astrophysics, Cambridge MA, USA}\label{cfa}
\and INAF - Osservatorio Astronomico di Roma, Monte Porzio Catone, Italy
\label{oar}
\and Institut de RadioAstronomie Millim\'etrique, Grenoble - France
\label{iram}
\and Leiden Observatory, Leiden University, Leiden, The Netherlands
\label{leiden}
\and Department of Physics and Astronomy, University College London, London, UK
\label{ucl}
\and Dept.~of Physics and Astronomy, University of Waterloo, Canada \label{waterloo} 
\and European Space Astronomy Centre, ESA, Madrid, Spain \label{esac}
}


\date{Received May 31, 2010 / Accepted July 7, 2010}

\abstract
{
In contrast to the more extensively studied dense star-forming cores, little is known about diffuse gas surrounding star-forming regions. 
}
{
We study the molecular gas in the Galactic high-mass star-forming region NGC\,6334I, which contains diffuse, quiescent components that are inconspicuous in widely used molecular tracers such as CO.
}
{
We present \textit{Herschel}/HIFI observations of methylidyne (CH) toward NGC\,6334I observed as part of the ``Chemical \textit{HErschel} Surveys of Star forming regions" (CHESS) key program. HIFI resolves each of the six hyperfine components of the lowest rotational transition ($J$=$\frac{3}{2}$--$\frac{1}{2}$) of CH, observed in both emission and absorption.
}
{
The CH emission features appear close to the systemic velocity of NGC\,6334I, while its measured FWHM linewidth of 3\,\kms\ is smaller than previously observed in dense gas tracers such as NH$_3$ and SiO. The CH abundance in the hot core is $\sim$\pow{7}{-11}, two to three orders of magnitude lower than in diffuse clouds. While other studies find distinct outflows in, e.g., CO and H$_2$O toward NGC\,6334I, we do not detect any outflow signatures in CH.
At least two redshifted components of cold absorbing material must be present at $-3.0$ and $+6.5$\,\kms\ to explain the absorption signatures. We derive a CH column density ($N_\mathrm{CH}$) of \pow{7}{13} and \pow{3}{13}\,\psqcm\ for these two absorbing clouds. We find evidence of two additional absorbing clouds at $+8.0$ and $0.0$\,\kms, both with $N_\mathrm{CH}$ $\approx$\pow{2}{13}\,\psqcm. Turbulent linewidths for the four absorption components vary between 1.5 and 5.0\,\kms\ in FWHM. We constrain the physical properties and locations of the clouds by matching our CH absorbers with the absorption signatures seen in other molecular tracers.
}
{
In the hot core, molecules such as H$_2$O and CO trace gas that is heated and dynamically influenced by outflow activity, whereas the CH molecule traces more quiescent material. 
The four CH absorbing clouds have column densities and turbulent properties that are consistent with those of diffuse clouds: two are located in the direct surroundings of NGC\,6334, and two are unrelated foreground clouds. Local density and dynamical effects influence the chemical composition of the physical components of NGC\,6334, which causes some components to be seen in CH but not in other tracers, and vice versa. 
}

\maketitle

\section{Introduction}
\label{sec:intro}

Most studies of star-forming regions focus on dense gas in the cores, while much less is known about surrounding diffuse clouds.
Methylidyne (CH) can be used as a tracer of low density clouds, since its column density is known to correlate with the total molecular column density in diffuse clouds \citep{vandishoeck1986,liszt2002,sheffer2008,chastain2010}. In contrast, most carbon is locked in CO in higher density environments ($n(\mathrm{H}_2) \gtrsim10^4$\,\pccm), resulting in a lower fractional abundance of CH \citep[e.g.,][]{polehampton2005}. 

Interstellar CH was first observed in its electronic transitions in the visual \citep{adams1941}, and later the \mbox{$\Lambda$-doublet} at 3.3\,GHz \citep{rydbeck1973,turner1974}. The accurate determination of a CH abundance from its radio lines is complicated by population inversion effects, while its pure rotational transitions lie in the far-infrared and are unobservable from the ground. While the rotational spectrum of CH has been studied since the 1970s \citep[see][]{davidson2001}, the first astronomical detection of a rotational transition of CH (at 149\,\micron) was reported by \citet{stacey1987} in Sgr~B2. 
With the Heterodyne Instrument for the Far-Infrared \citep[HIFI,][]{degraauw2010} onboard the \textit{Herschel} Space Observatory \citep{pilbratt2010}, it is now possible to observe the ground-state rotational transition of CH at 533 and 537\,GHz, free from atmospheric interference and with unprecedented spectral resolution.

In this Letter, we study spectrally resolved signatures of the CH $^2\Pi_{3/2}$ $J$=3/2--1/2 transition, with an upper level energy of 26\,K, toward NCG\,6334I, which exhibits emission from the hot core itself, and absorption due to cold foreground clouds. 
NGC\,6334I is a high-mass star-forming core at a distance of 1.7\,kpc belonging to the luminous NGC\,6334 complex of molecular clouds. The ``I" source is composed of several dense subcores \citep{hunter2006} with a total mass of $\sim$$200\,M_\odot$ and a bolometric luminosity of several $10^5\,L_\odot$ \citep{sandell2000}. It is rich in molecular lines \citep[e.g.,][]{schilke2006}. Several foreground clouds have been identified along the line of sight toward NGC\,6334 by \citet{brooks2001} and \citet{beuther2005}. 
We use \textit{Herschel}/HIFI data to estimate the CH abundance in the NGC\,6334I hot core and to constrain the physical properties of the CH absorbing foreground clouds.

\section{Observations and data reduction}
\label{sec:obs}

We use data from a spectral scan from the ``Chemical \textit{HErschel} Surveys of Star forming regions" key program (CHESS; \citealt{ceccarelli2010}), observed by HIFI on board the \textit{Herschel} Space Observatory. The band 1a scan was performed on February~28, 2010 using the dual beam switch (DBS) mode with continuum stability optimization. The half-power beamwidth at the relevant frequencies is about 42\arcsec\ and was centered at $\alpha_{J2000}$=$17^{\rm h}20^{\rm m}53\fs32$, $\delta_{J2000}$=$-35\degr46\arcmin58.5\arcsec$, which is between the SMA1 and SMA2 cores in Fig.~1 of \citet{hunter2006}. The reference positions for the DBS were $\sim$3\arcmin\ east and west of the target. 
HIFI spectral scans use the wideband spectrometer (WBS) with a frequency resolution of 1.1\,MHz, corresponding to 0.6\,\kms\ at the frequencies of the two groups of CH lines. The full band 1a (480--560\,GHz) scan took 12\,950 seconds, including overheads. The resulting noise level is 12\,mK in \Tmb\ units, measured in 0.5\,MHz channels, in the `single sideband' spectrum (see below), averaged between H- and V-polarization. The double sideband system temperature varied between 60 and 110\,K.

The raw data is pipelined in HIPE~2.6 \citep{ott2010}. We note that the absolute velocity scale is uncertain because of improper calculation of the spacecraft motion, but this has been constrained to 0.3\,\kms\ by subsequent reprocessing in a more recent pipeline version. The resulting level~2 product was processed in CLASS\footnote{Version of April 2010, with some improvements. CLASS is part of the GILDAS software developed at IRAM: \texttt{http://www.iram.fr/IRAMFR/GILDAS}.}. After mixing the sky signal with a local oscillator (LO) signal, a heterodyne detector such as HIFI yields a double sideband spectrum. In spectral scans, each frequency channel is observed using several LO settings (observations discussed here use a redundancy of 8), which makes it possible to mathematically disentangle the signal from the two separate sidebands \citep{comito2002}. This `deconvolution' step is executed in CLASS, while allowing channel-to-channel variations in receiver gain. Resulting gain variations are 10\% at maximum. Spectra presented in this Letter are always deconvolved single sideband spectra. 
Intensities are corrected for a main beam efficiency of 70.8\% expected for band 1a of HIFI.

\section{Line profile analysis}
\label{sec:results}

\begin{figure}
  \centering
  \resizebox{\hsize}{!}{\includegraphics{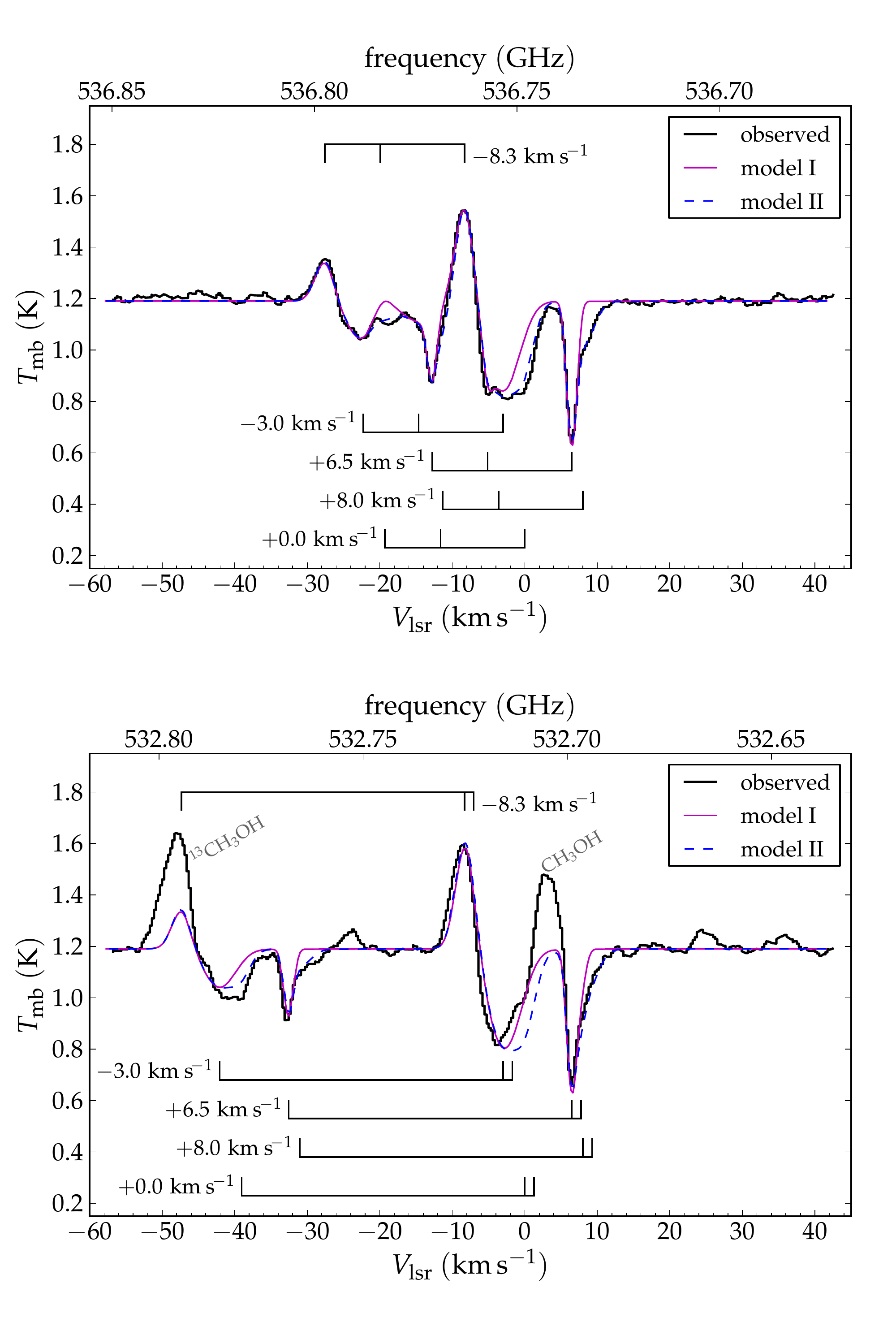}}
  \caption{Observed spectra (black) of the two groups of three hyperfine components of the CH $^2\Pi_{3/2}$ $J$=3/2--1/2 transition, represented in \Tmb\ units along the vertical axis (see Sect.~\ref{sec:obs}). The horizontal axis gives frequency (top) and velocity with respect to the rest frequency for the strongest line in each triplet (bottom). The magenta line represents model I with one emission component and two absorption components and the blue line represents model II with two additional absorption components. Physical parameters of the model components are given in Table~\ref{t:modelparam}. `Comb' labels indicate the loci of the CH hyperfine lines shifted with respect to the rest frame frequency.} 
  
  \label{fig:CHspectra}
\end{figure}

The single sideband HIFI WBS spectra near 536.76 and 532.73\,GHz (Fig.~\ref{fig:CHspectra}) show continuum emission with \Tmb=1.19\,K and several spectral lines. The two spectral segments, each about 0.2\,GHz wide, cover the CH $^2\Pi_{3/2}$ $J$=3/2--1/2 parity $^+$ to $^-$ transitions around 532.73\,GHz and the parity $^-$ to $^+$ transitions around 536.76\,GHz \citep{davidson2001}. Most lines are due to CH, but two emission features in the 532.73\,GHz spectrum have contributions from the \methanol\ $J_K$=$15_2$--$15_1$ and the $^{13}$\methanol\ $9_3$--$9_2$ transitions.

\subsection{Model setup}
\label{sec:modelsetup}

\begin{table} 
\caption{Model component parameters.}
\label{t:modelparam} 
\begin{tabular}{l c c c c c} 
\hline\hline 
\noalign{\smallskip}
Component	& \Tex 	& \vlsr	& FWHM	& \multicolumn{2}{c}{$N_\mathrm{CH}$ ($10^{13}$ \psqcm)}	 \\
			& (K)		& (\kms)	& (\kms) 	&  model I & model II    \\
\hline
Emission		& 100	& $-8.3$	& 3.0	& 20       & 20 \\
Absorption A	& 5.5	& $-3.0$ 	& 5.0	& 6.8      & 6.6 \\
Absorption B	& 5.5	& $+6.5$	& 1.5	& 3.8      & 3.0 \\	
Absorption C    & 5.5	      & $+8.0$	 & 3.5	    & \ldots  & 2.1 \\
Absorption D	   & 5.5		& $+0.0$ & 3.0        &  \ldots  & 2.1 \\
\hline
\end{tabular} 

\end{table}

From inspection of the observed spectrum in Fig.~\ref{fig:CHspectra}, we note that the two sets of three emission peaks are spaced by exactly the hyperfine splitting. Since we see more than three absorption features near each emission triplet, we conclude that more than one absorption component is necessary to explain the observed spectra. We therefore generate a model CH spectrum starting from the minimum of one emitting and two absorbing components (`model I', see Table \ref{t:modelparam}) using the CASSIS software\footnote{CASSIS has been developed by CESR-UPS/CNRS (http://cassis.cesr.fr).} and the CDMS spectroscopic database \citep{muller2005}. Given an excitation temperature and a column density for a set of emission and absorption lines, the `LTE' module in CASSIS calculates a frequency-dependent optical depth, which is used to construct a beam-averaged, frequency-dependent line brightness.

We adopt a source size of 10\arcsec\ for the emission component, taken to encompass the four SMA cores \citep{hunter2006}. With a beam size of 42\arcsec\ for HIFI band 1a, this results in a beam-filling factor of 6\%. Since we lack collisional rate coefficients for CH, we take the dust temperature of 100\,K from \citet{sandell2000} and assume this to be equal to the excitation temperature (\Tex) of the emitting CH gas in NGC\,6334I. We note that, while we assume a single value for \Tex, the nominal dust temperatures determined from interferometric observations by \citet{hunter2006} vary between 33 and 100\,K from core to core, with limits of 20 and 300\,K. The continuum level for the model is fixed at the observed level of 1.19\,K. 
For the absorbing foreground components, \Tex=5.5\,K is based on H$_2$O absorption signatures discussed in \citet{emprechtinger2010}.

\subsection{Column density estimates}

Since all features in the 536.76\,GHz spectrum (Fig.~\ref{fig:CHspectra}, top panel) can be ascribed to CH, we use this as a starting point to constrain the remaining model parameters for the three components: line of sight velocity (\vlsr), column density ($N_\mathrm{CH}$), and full-width at half-maximum of the line (FWHM). The best-fit set of parameters presented in Table~\ref{t:modelparam} (model~I) are: emission at \vlsr=$-8.3$\,\kms\ with FWHM=3.0\,\kms\ and $N_\mathrm{CH}$=\pow{2.0}{14}\,\psqcm\ and two absorbing components at $-3.0$ and $+6.5$\,\kms, with distinctly different line widths of 5.0 and 1.5\,\kms, respectively. 
After fine-tuning the parameters of the three components, some discrepancies between model I and the observed spectra remain. Extra absorption appears to be required near 536.782, 536.748 and 536.732\,GHz (see Fig.~\ref{fig:CHspectra}, top panel). The frequency spacing between these discrepant regions is such that they cannot be explained by a single additional component.

The discrepancies can be resolved by adding two absorbing clouds, both with \Tex=5.5\,K and $N_\mathrm{CH}$=\pow{2}{13}: component C with \vlsr=+8.0\,\kms\ and FWHM=3.5\,\kms, and component D with \vlsr=0.0\,\kms\ and FWHM=3.0\,\kms\ (Fig.~\ref{fig:CHspectra}). Upon introducing components C and D, the column densities of the absorbing components A and B need to be slightly decreased to \pow{6.6}{13} and \pow{3.0}{13}\,\psqcm, respectively (`model II', Table~\ref{t:modelparam}).

Without specific tailoring of the model to match the 532.73\,GHz spectrum, it explains this second set of CH signatures remarkably well, with the exception of the contamination by \methanol\ and $^{13}$\methanol\ (see Fig.~\ref{fig:CHspectra}, bottom panel). 
Although the addition of components C and D moves the model closer to the spectrum of the 536.76\,GHz triplet, the correspondence to the 532.73\,GHz spectrum still has some notable discrepancies: the absorption feature near 532.78\,GHz is not deep enough, and the absorption signature near 532.72\,GHz appears to be shifted by $\sim$2\,\kms. The latter mismatch could be alleviated by adding a methanol emission line. Additional contamination by other species is not ruled out at the level of $\sim$0.05\,K line intensity, as illustrated by emission features around both 532.670 and 532.754\,GHz, and possibly the blue wings of the methanol lines.

\subsection{Uncertainties}

Variations in column density for an individual model component have a roughly linear effect on the equivalent width of the corresponding spectral feature, i.e., doubling $N_\mathrm{CH}$ (dividing by two) leads to an emission or absorption line that is twice as strong (weak). 
Moreover, the column densities that we derive for each component depend directly on the assumed values of \Tex. We briefly explore the effect of changing \Tex\ on the line intensities, which we compensate by adjusting the column density. For the emitting gas, raising \Tex\ from 100 to 150\,K requires a compensating column density increase of $\sim$35\%, while lowering \Tex\ from 100 to 50 (to 30)\,K requires a column density decrease of only 20\% (15\%). 
Because the CH transition under consideration becomes optically thick at low \Tex, the dependence of column density on \Tex\ weakens and eventually reverses as \Tex\ is lowered to $\sim$50\,K and beyond. To illustrate, if \Tex\ is varied from 150 to 100 to 50 to 30\,K, the opacity for the main emission line increases from 0.05 to 0.07 to 0.19 to 0.39, respectively. The line opacity therefore reacts more strongly than linearly to a change in \Tex, resulting in counterintuitive behavior of line strength as a function of \Tex. 
For the cold absorbing gas, an increase in \Tex\ from 5.5 to 20\,K roughly doubles the column density, and changing \Tex\ from 5.5 to 2.7\,K results in a column density decrease of $<$10\%. This illustrates a much weaker dependence of column density on \Tex\ than for the emitting gas. Optical depths reach maxima of 0.34, 0.51, 0.15, and 0.18 at the respective line centers for absorbing components A, B, C, and D. 

The uncertainty in our derived column density for the emission component is dominated by the uncertainty in the \Tex\ estimates ($\lesssim$\,2, see Sect.~\ref{sec:modelsetup}). Following the above discussion, quoted column density values can be trusted taking into account an error margin of $\sim$30\% for the warm emitting gas. On the other hand, the uncertainty in column densities for the cold absorbing gas components is dominated by measurement errors in the continuum and absorption line depth, each estimated at 6\%, resulting in an uncertainty of 8\% for the column densities.

\section{Discussion}
\label{sec:discussion}

\subsection{CH in the hot core}

The velocity of the emission component at $-8.3$\,\kms\ is consistent with the H$_2$O emission detected by \citet{emprechtinger2010} at $-8.1$\,\kms\ and close to the velocity of $\approx -7$\,\kms\ of molecular lines measured from the ground \citep{nummelin1998}. The width of the CH emission features of 3.0\,\kms\ is smaller than the 4--5\,\kms\ for NH$_3$ \citep[ATCA at 24\,GHz;][]{beuther2005}, 6.5\,\kms\ for SiO, and 6\,\kms\ for HC$_3$N \citep[IRAM 30m at 87, 130 and 155\,GHz;][]{bachiller1990}. This difference suggests that the CH emission originates in different, less turbulent and/or less evolved, cores than the heavier N- and O-bearing molecules. \citet{lis2010a} find a similarly narrow linewidth of 3.3\,\kms\ from HIFI observations of H$^{37}$Cl 1--0 emission from the NGC\,6334I hot core. 
\citet{sandell2000} derives a source size of (10$\times$8)\arcsec\ and a dust temperature of 100\,K from sub-millimeter dust continuum measurements. Combining this with a 350\,\micron\ flux from SHARC-II, dust grain opacities from \citet{ossenkopf1994} and the distance of 1.7\,kpc, we arrive at $N_\mathrm{H_2}$$\approx$\pow{3}{24}\,\psqcm. From this $N_\mathrm{H_2}$ and our $N_\mathrm{CH}$=\pow{2.0}{14}\,\psqcm, we find an abundance for CH of \pow{7}{-11} relative to H$_2$ in the hot core. This is a factor $\sim$7--30 lower than the CH abundance for the dense envelope of Sgr~B2, and a factor $\sim$300 lower than observed CH abundances in diffuse clouds \citep[][Table~4, and references therein]{polehampton2005}. This decrease in CH abundance is similar to that predicted by chemical models (Sect.~\ref{sec:intro}) for H$_2$ densities of $\gtrsim$$10^4$\,\pccm.

\subsection{CH in absorbing clouds}
\label{sec:disc_abs}

The range of turbulent linewidths for our four absorbing clouds (Table~\ref{t:modelparam}) and $N_\mathrm{CH}$ values of several $10^{13}$\,\psqcm\ agree with models of `turbulence dissipation regions' \citep{godard2009} and with optical observations of diffuse clouds \citep{liszt2002}.
Using the range of $N_\mathrm{CH}$ values for absorption clouds A--D (Table~\ref{t:modelparam}) and an $N_\mathrm{CH}/N_\mathrm{H_2}$ ratio of \pow{3.5}{-8} \citep{sheffer2008}, we find $N_\mathrm{H_2}$=\pow{(0.6--2)}{21}\,\psqcm. Dividing $N_\mathrm{H_2}$ of $10^{21}$\,\psqcm\ by a number density of $10^2$--$10^3$\,\pccm\ applicable to diffuse gas, we derive a rough linear size estimate of 0.3--3\,pc for the absorbing clouds. We also convert $N_\mathrm{H_2}$ =\pow{(0.6--2)}{21}\,\psqcm\ into a visual extinction, $A_V \sim 0.3$--$1$, based on $N_\mathrm{H}$/$A_V$=\pow{1.87}{21}\,\psqcm\,mag$^{-1}$ \citep{bohlin1978}. Hence, we conclude that the CH absorption components are diffuse interstellar clouds. 
We note that, although we discuss several absorbing components as separate entities in velocity space, a diffuse interstellar cloud of several parsecs in size can enjoy a velocity gradient of a few \kms\ (cf.~FWHM in Table~\ref{t:modelparam}) and therefore various components may originate in a single molecular cloud.

Other studies also find absorbing gas along the line of sight toward NGC\,6334I. We summarize those that potentially correspond to our CH clouds. \citet{brooks2001} note two clouds absorbing in OH (near 1.6\,GHz) toward several NGC\,6334 cores: one cloud with velocities ranging between $-15$ and $+2$\,\kms, which they assume to be associated with the NGC\,6334 complex, and another at $+6$\,\kms\ which they ascribe to a more extended foreground cloud. In addition, the HIFI study of H$_2$O lines toward NGC\,6334I by \citet{emprechtinger2010} reports absorbing foreground clouds at $-6.3$, $-0.1$ and $+6.1$\,\kms. 

In terms of \vlsr, our absorption components A and D (Table~\ref{t:modelparam}) fit in the widespread $[-15, +2]$\,\kms\ OH absorption clouds, which places them physically close to the NGC\,6334 molecular cloud. In addition, cloud D coincides with one of the three H$_2$O absorbing clouds. Hence, although $N_\mathrm{CH}$ for cloud A is three times higher than for cloud D, we conclude that cloud D must be of somewhat higher \emph{number} density to allow for H$_2$O formation.
Moreover, \citet{ossenkopf2010_H2O+} find H$_2$O$^+$ absorption toward NGC\,6334I. There is notable uncertainty regarding the rest frequencies of the H$_2$O$^+$ ground state rotational transitions. We assume a value of 1115.209\,GHz for $F$=5/2--3/2, which is close to predictions by \citet{muertz1998}, to derive \vlsr$\approx$$-1$\,\kms\ for the H$_2$O$^+$ absorption. Both the above H$_2$O$^+$ signature and the detection of H$_2$Cl$^+$ absorption at $-1.7$\,\kms\ with a width of 12\,\kms\ \citep{lis2010a} are consistent with a blend of our CH components A and D in velocity space.

Our component B at $+6.5$\,\kms\ is consistent with both the $+6.1$\,\kms\ H$_2$O absorption \citep{emprechtinger2010} and the $+6$\,\kms\ OH component to within the 0.7\,\kms\ velocity resolution of the ATCA observations \citep{brooks2001}. We believe that this cloud has a relatively high particle density, but is unrelated to the NGC\,6334 complex.
Cloud C is seen in CH but has no counterparts in H$_2$O or OH. The column density of this cloud is not particularly low with respect to clouds A, B, and D (cf.~Table~\ref{t:modelparam}), but it might simply be of lower number density. In conclusion, from the correspondence with counterparts in other molecules, we find that CH clouds B and C are unrelated foreground clouds, while clouds A and D are located in the direct surroundings of NGC\,6334I.

\subsection{Dense clouds and outflows}
\label{sec:disc_noCH}

In addition to the absorbing clouds discussed in Sect.~\ref{sec:disc_abs}, the literature provides a collection of absorption signatures without CH counterparts.
First, we find no evidence of absorbing CH gas that matches the $-6.3$\,\kms\ component seen in H$_2$O \citep{emprechtinger2010}. The low CH abundance in this cloud -- close to the NGC\,6334I hot core in velocity space -- may be due to its physical proximity to the luminous NGC\,6334 region, where higher gas densities facilitate active chemistry, leading to depletion of CH. In general, overall column density and visual extinction are traced more accurately by CH than by H$_2$O and CO, which are very sensitive to inhomogeneity \citep[e.g.,][]{spaans1996}. 

Second, \citet{beuther2005} detect \methanol\ and NH$_3$ absorption at 25\,GHz toward NGC\,6334-I-SMA3 (the compact \HII\ region) at $-9$\,\kms, about 2\,\kms\ blueward of the emission component. 
Moreover, outflows with velocities up to 50\,\kms\ are observed in H$_2$O \mbox{$1_{11}$--$0_{00}$}, \mbox{$2_{11}$--$2_{02}$}, \mbox{$1_{10}$--$1_{01}$} and \mbox{$3_{12}$--$3_{03}$} emission \citep{emprechtinger2010} and up to 70\,\kms\ in CO $J$=\mbox{2--1}, \mbox{3--2}, \mbox{4--3} and \mbox{7--6} \citep{bachiller1990,leurini2006b}. 
In our HIFI observations of the CH transition, we detect neither blueshifted gas nor outflows to match the H$_2$O, \methanol, NH$_3$, CO and H$_2$O$^+$, in neither emission nor absorption. The absence of CH in both blueshifted emission and outflows indicates that CH is destroyed in shocked environments. We conclude that CH traces molecular material that has not yet been swept up by outflow activity related to star formation.

\section{Conclusions}

We use CH as a tracer of cold, quiescent, molecular gas in both the star-forming core of NGC\,6334I and diffuse clouds along the line of sight. We derive a very low CH abundance of \pow{7}{-11} in the hot core, consistent with CH depletion in chemically active dense cores. Evidence is found of four absorbing components composed of diffuse gas. From comparison with other molecular absorption signatures in the literature, we conclude that clouds A and D are related to the NGC\,6334 complex, and that clouds B and C are unrelated foreground clouds (Table~\ref{t:modelparam}). Differences between the four clouds in terms of counterparts in other molecular tracers are ascribed to local variations in density and dynamics. Dynamical stirring of the gas is hypothesized to be the cause of the absence of CH in the outflow component, which \emph{is} detected in other molecular tracers.

\begin{acknowledgements}
HIFI has been designed and built by a consortium of institutes and university departments from across Europe, Canada and the United States under the leadership of SRON Netherlands Institute for Space Research, Groningen, The Netherlands and with major contributions from Germany, France and the US. Consortium members are: Canada: CSA, U.\,Waterloo; France: CESR, LAB, LERMA, IRAM; Germany: KOSMA, MPIfR, MPS; Ireland, NUI Maynooth; Italy: ASI, IFSI-INAF, Osservatorio Astrofisico di Arcetri-INAF; Netherlands: SRON, TUD; Poland: CAMK, CBK; Spain: Observatorio Astron\'omico Nacional (IGN), Centro de Astrobiolog\'ia (CSIC-INTA). Sweden: Chalmers University of Technology - MC2, RSS \& GARD; Onsala Space Observatory; Swedish National Space Board, Stockholm University - Stockholm Observatory; Switzerland: ETH Z\"urich, FHNW; USA: Caltech, JPL, NHSC. 
We thank Marco Spaans for useful discussions, and the referee Volker Ossenkopf for constructive suggestions. 
\end{acknowledgements}

\bibliographystyle{aa}  
\bibliography{../../literature/allreferences}

\end{document}